\documentstyle[preprint,aps,eqsecnum]{revtex}
\begin{document}
\draft
\preprint{}

\tightenlines

\title   {
Tetrahedral and Triangular Deformations of $Z=N$ Nuclei \\
in Mass Region $A \sim 60-80$
         }


\author  {
S. Takami, K. Yabana and M. Matsuo$\hbox{}^{*}$
         }
\address {
Graduate School of Science and Technology,
Niigata University, Niigata 950-21 \\
$\hbox{}^*$Yukawa Institute for Theoretical Physics, Kyoto University,
Kyoto 606-01
         }

\maketitle

\vspace{15mm}

\begin{abstract}
We study static non-axial octupole deformations in proton-rich $Z=N$
nuclei, 
$^{64}$Ge, $^{68}$Se, $^{72}$Kr, $^{76}$Sr, $^{80}$Zr and $^{84}$Mo, 
by using the Skyrme Hartree-Fock plus BCS calculation with no
restrictions on the nuclear shape.
The calculation predicts that the oblate ground state in $^{68}$Se
 is extremely soft for the $Y_{33}$ triangular deformation, and that
 in $^{80}$Zr the low-lying local minimum state coexisting 
with the prolate ground state  has the $Y_{32}$ tetrahedral deformation.
\end{abstract}


\vspace{15mm}


\newpage

Spontaneous intrinsic deformation in the ground and excited 
states is one of the most fundamental properties of nuclei.
In addition to the well-established  quadrupole deformation,
octupole deformation violating reflection symmetries has been 
attracted much experimental and theoretical attentions.
Recently, a pear-shape deformation, i.e., an axially symmetric $Y_{30}$
octupole deformation superposed on a prolate quadrupole deformation has 
been established in light actinide
(Ra-Th isotopes around $A \sim 220$) and lanthanide
(Xe-Ba isotopes around $A \sim 136$) mass regions\cite{BN96}. 
More exotic shapes such that {\it violate
both reflection and axial symmetries} are known 
in light $p$ and $sd$-shell nuclei,
where the exotic shapes
are caused by $\alpha$ cluster correlations
\cite{CLUSTER},
like as
the equilateral triangular 3$\alpha$ configuration in $^{12}$C.
Besides cluster models assuming such configuration,
mean-field models\cite{Eichler,Leander,TETRA} 
have also been discussed them in terms of
non-axial octupole deformations. 
It is of great interest to see whether the non-axial exotic octupole
deformations realize in heavier systems.
Possibilities
of the non-axial static octupole deformations have been argued theoretically
for superdeformed nuclei\cite{Li-SD,Chasman} and for the
light Ra region\cite{LD94}, for which however 
there seem to be no experimental support so far.

Proton-rich nuclei in the $A \sim 60 - 80$ mass region are advantageous 
candidates for
a search of the exotic reflection asymmetric deformations.
For nuclei with neutron and proton numbers around 30-40, 
octupole correlation  can be expected 
because of $\Delta j=3$ coupling between the $2p_{3/2}$ and 
$1g_{9/2}$ orbitals in the major $pfg$-shell\cite{Nea84,Lea82,BN96}. 
Furthermore, the shell effect which is the important origin 
of the deformations is generally large in 
comparison with those in light lanthanide and actinide nuclei.
Especially in proton-rich $Z=N$ nuclei, both proton and neutron 
configurations 
coherently operate to develop a static octupole deformation
although there is no available experimental information such as 
$E_{x}(3^{-}_{1})$ and B(E3) in proton-rich $Z=N$ nuclei 
around $A \sim 80$, except for $^{64}$Ge \cite{Eea91}.
The previous theoretical studies based on the shell correction 
method suggested that the octupole instability is not
strong enough \cite{Nea84,Eea91,Skalski} to develop 
a static octupole deformation in this mass region.
However, it should be noted that these models considered only the
axially symmetric deformation \cite{Nea84,Eea91}, 
or limited combination of 
non-axial deformation parameters \cite{Skalski}.

In this letter, we will show that the reflection asymmetric shapes 
violating  axial symmetry are more favored in $Z=N$ nuclei 
in the mass $A=60-80$ region 
than that with axial symmetry.
We perform a fully self-consistent mean-field calculation
for even-even
nuclei $^{64}$Ge, $^{68}$Se, $^{72}$Kr, $^{76}$Sr, $^{80}$Zr and
$^{84}$Mo by means of 
the Skyrme Hartree-Fock plus BCS method with three-dimensional (3D)
mesh representation\cite{Bea85}.
In contrast to the previous calculations with 3D mesh representation, 
we impose no requirement of symmetry on the solutions to allow
arbitrary nuclear shapes.

In the present method,  quantities such as
single-particle wave functions and densities
are described 
on a 3D Cartesian mesh within spherical box.
The radius of spherical box and the width of the 3D mesh 
are set to 
12 fm and 1fm, respectively.
Imposing the constraints which diagonalize the mass inertia
tensor, we make the principal axes coincide
with $x$, $y$ and $z$ axes of the mesh.
We adopt the Skyrme III interaction, which has been successful in 
describing systematically the ground state
quadrupole deformation in proton and neutron rich Kr,Sr,Zr and Mo
isotopes \cite{Bea85} and in a wide area of nuclear chart \cite{TTO96}.
It is also able to describe the axially symmetric octupole deformation 
in the Ra-Th region \cite{Bea86}.
As for the pairing strength of proton, we use the same 
parameterization $G_{p}=16.5/(11+Z)$ MeV as in Ref. \cite{Bea85}
together with the same truncation of the single-particle space. 
The neutron pairing strength is taken the same as $G_{p}$ \cite{Heenen}.
To characterize deformation of the obtained solutions, we have calculated 
the mass multipole moments,
\begin{equation}
	\alpha_{lm} \equiv \frac{4\pi \langle \Phi | 
		\sum_{i}^{A} r_{i}^{l} X_{lm}(i) | \Phi \rangle}
		{3 A R^{l}}, (m=-l,\cdots,l),
\end{equation}
where $A$ is the number of nucleon and $R = 1.2 A^{1/3}$ fm.
Here $X_{lm}$ is a  real basis of the spherical harmonics,
\begin{eqnarray}
	X_{l0} & = & Y_{l0},
		\nonumber \\
	X_{l|m|} & = & \frac{1}{\sqrt{2}}( Y_{l-|m|}+Y_{l-|m|}^{*} ),
		\nonumber \\
	X_{l-|m|} & = & \frac{-i}{\sqrt{2}} ( Y_{l|m|}-Y_{l|m|}^{*} ),
\end{eqnarray}
where the quantization axis is chosen as  the largest and smallest
principal inertia axes for  prolate and oblate solutions, respectively.
For the quadruple moment, we also use ordinary
($\beta,\gamma$) notation, i.e. , $\alpha_{20}=\beta\cos{\gamma},
\sqrt{2}\alpha_{22}=\beta\sin{\gamma}$, mapped in the $\beta>0, 0<\gamma<
\pi/3$
sections.
To represent magnitude of the octupole deformation, we define
\begin{equation}
	\beta_{3}\equiv ( \sum_{m=-3}^{3} \alpha_{3m}^{2} )^{\frac{1}{2}},
	\; \; \; \; \; \;
	\beta_{3m} \equiv ( \alpha_{3m}^{2}+\alpha_{3-m}^{2} )^{\frac{1}{2}}
	\; \; \; \; \; \;
	  (m=0,1,2,3).
\end{equation}
For nuclei around $A \sim 80$, existence of three local energy 
minimum states showing oblate, nearly spherical and prolate 
solutions has been reported in the SHF+BCS calculations\cite{Bea85,TTO96}.
To search all minimum states close energetically 
to the ground state, we generate initial states by solving 
a deformed Wood-Saxon potential model.
The five initial states with different quadrupole 
deformations are used : 
(1) $\beta = 0.7, \gamma = 60^{\circ}$, 
(2) $\beta = 0.3, \gamma = 60^{\circ}$, 
(3) $\beta = 0.0, \gamma = 0^{\circ}$, 
(4) $\beta = 0.3, \gamma = 0^{\circ}$, 
(5) $\beta = 0.7, \gamma = 0^{\circ}$.
For all initial configurations, the octupole distortion,
 $\alpha_{3 m}=0.1$ ($m = -3,\cdots,3$), is added.

In table \ref{table1}, we summarize the
quadrupole and octupole deformation parameters of the
obtained solutions and the excitation energy of the local
minimum states,
 where we do not list states higher than the fourth minimum.
The solutions are classified into three groups, oblate, 
spherical and prolate, according to their quadrupole deformations. 
As is seen in table \ref{table1}, the calculated quadrupole 
deformation of the ground states varies significantly as 
changing the mass number.
For $Z=N=32$ ($^{64}$Ge), the calculation results in 
the triaxial deformation, which is in agreement 
to the experimental indications\cite{Eea91}.
It is also seen that the ground state quadrupole deformation
changes suddenly from the moderate oblate ones in $^{68}$Se and $^{72}$Kr
to the strong prolate ones in  $^{76}$Sr and $^{80}$Zr.
Sudden change of the quadrupole deformation qualitatively 
agrees with the observed systematics of the first excited energy 
levels (considered  to be $J^{\pi}=2^{+}$ state)
from $^{64}$Ge to $^{84}$Mo \cite{Lea90-Gea91}.
For all the calculated elements, we find the 
low-lying local  minimum states with different quadrupole deformations,
which is a characteristic of shape coexistence nuclei.

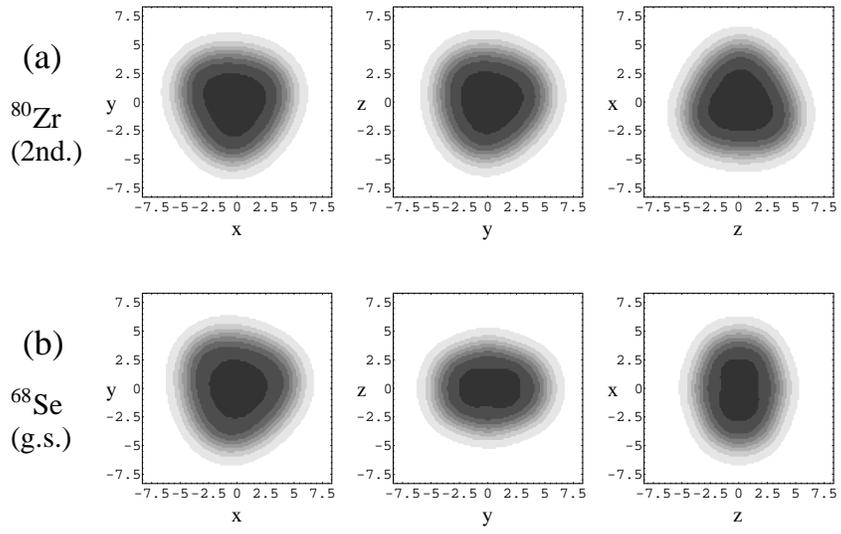
\begin{figure}
\caption{Density distributions of proton in the $xy$, $yz$ and $zx$ 
planes where $x$, $y$ and $z$ axes represent the principal inertia
axes. 
(a) and (b) show those of the second minimum state of $^{80}
$Zr and the ground state of $^{68}$Se, respectively.}
\label{figure1}
\end{figure}

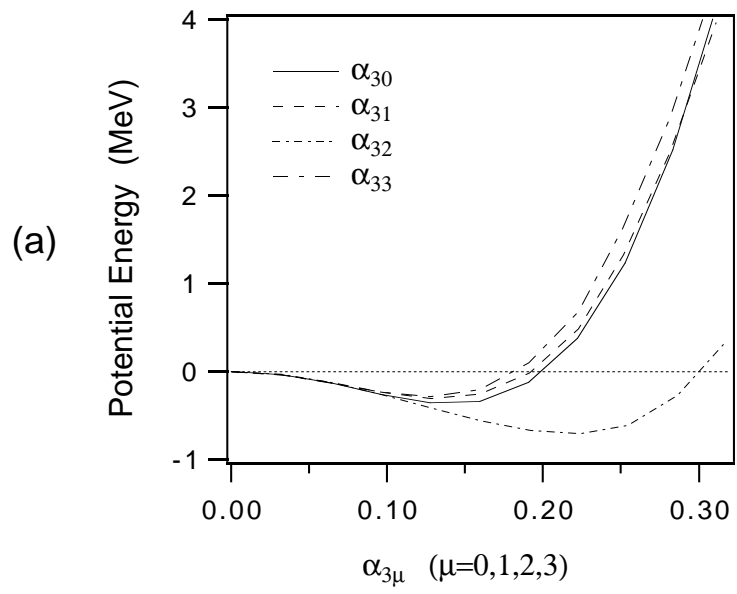
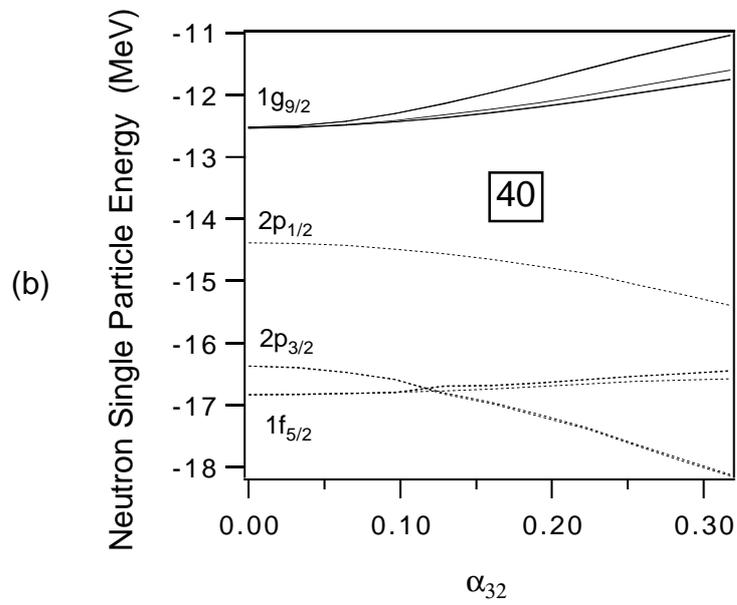
\begin{figure}
\caption{(a) The potential 
energy surfaces of $^{80}$Zr with respect to the different types 
of the octupole 
deformations, where the energy is 
measured in relative to the spherical solution.
The potential energy is calculated as a function of  $\alpha_{3m}$ 
$(m=0,1,2,3)$ by imposing the constraints of 
$\beta=0, \gamma=0^{\circ}$ and $\alpha_{3\nu} =0$ 
( $\nu \neq m$ ).
(b) The single particle energy of neutron as a function of the
tetrahedral $\alpha_{32}$ deformation.}
\label{figure2}
\end{figure}

The octupole deformations violating the axial symmetry are 
found in the ground state or local minimum states in
all nuclei, except $^{64}$Ge (where the obtained $\beta_3=0.01$ is not
sizable). 
Among them, the second minimum state of $^{80}$Zr shows 
the largest octupole deformation of $\beta_{3}=\beta_{32}=0.24$ 
without having a quadrupole deformation.
The density profile of this solution shown in Fig. \ref{figure1} (a)
indicates a tetrahedral deformation, which 
violates the both reflection and axial symmetries, but
obeys the symmetry of the point group $T_{d}$.
Figure \ref{figure2} (a) shows the 
potential energy surfaces of $^{80}$Zr with respect to 
the $\alpha_{30}$, $\alpha_{31}$, $\alpha_{32}$ and 
$\alpha_{33}$ deformations, which are calculated under the additional
 constraints of the mass octupole moments.
The potential energy surface of the $\alpha_{32}$ 
deformation has the minimum point at $\alpha_{32} = 0.24$ which
corresponds to the calculated lowest minimum, and 
the energy gain measured from the spherical solution
is as large as 0.71 MeV. 
This energy gain reduces the energy difference between the 
spherical and strongly prolate deformed solutions from 1.61 MeV to 0.90 MeV.
Octupole instability towards the $\alpha_{32}$ direction
(the tetrahedral deformation) is quite contrasting to 
 the other types ($\alpha_{3m} (|m|\neq 2)$)
of the octupole deformations.

Instability of the spherical configuration at $Z=N=40$ for the 
tetrahedral deformation can be ascribed to 
the shell effect formed in the potential having the $T_{d}$ symmetry.
In Fig.\ref{figure2} (b), we display 
the neutron single particle energies as a function of 
the tetrahedral deformation parameter $\alpha_{32}$.
As developing the tetrahedral $\alpha_{32}$ deformation of $^{80}$Zr,
 the orbitals 
stemming from $2p_{3/2}$ and $2p_{1/2}$ decrease in energy and those 
 stemming from $1g_{9/2}$ increase  with 
holding high degeneracy of orbitals.
The sub-shell gap at nucleon number 40 enhanced 
by  addition of $\alpha_{32}$ distortion field stabilizes 
the strongly tetrahedral deformed solution.
It is known that 
high degeneracy of irreducible representation of the $T_{d}$ 
symmetry tends to produce a significant bunch in the single 
particle level spectrum as has been demonstrated for electrons
in a metallic cluster potential by Hamamoto {\it et. al.}\cite{Hea91,FHM94}.
This tendency exists in nuclear potential with spin-orbit force.
Appearance of the tetrahedral deformation due to 
similar shell effect will not be confined in this neutron/proton
number as discussed by Li and Dudek for light actinide isotopes\cite{LD94}.
It is also noted that the tetrahedral octupole deformation
competes with
the pairing correlation which favors spherical deformation.
The energy gain caused by the tetrahedral deformation is 
found from calculation
to be more than 2 MeV when we neglect the pairing.
The tetrahedral minimum remains even with  30\% increase of the
paring strength ($G_{p},G_{n} \sim 0.4$ MeV).
It should be mentioned that the measured excitation energies of the
fist $3^-$ levels in Ge and Se isotopes have the minimum points
at $N=40$ \cite{Cottle},
which may be a fingerprint of octupole instability.

The other types of non-axial octupole deformation are also found.
The $^{68}$Se is the most noticeable
since it has the large octupole deformation ($\beta_3 = 0.15$)
in the ground state. 
As shown in the density distribution 
plotted in Fig.1(b), it has  $Y_{33}$ triangular
 distortion superposed 
on the oblate quadrupole deformation, which obeys the $D_{3h}$
symmetry seen in the regular triangular prism shape.
Although the minimum is not as deep as
the tetrahedral deformation in $^{80}$Zr, the potential
energy surface is quite flat up to $\alpha_{33} \sim 0.2$
as shown in Fig.3 (a).
It should be noted that octupole instability 
emerges only for the $\alpha_{33}$ direction.
Instability toward
the triangular deformation is systematically seen for the excited
oblate minima in  neighboring $^{72}$Kr and $^{76}$Sr.

Instability of the oblate states toward the triangular 
$Y_{33}$ deformation can also be related to the
single-particle  shell structure 
formed in the oblate deformed potential.
Figure \ref{figure4} shows the neutron Nilsson diagram as a
function of quadrupole deformation obtained in the constrained 
SHF+BCS method, in which axial and reflection symmetries are imposed.
In the oblate configuration of $^{68}$Se,
the $N,Z=34$ Fermi surfaces are located between 
the positive parity orbitals with 
$\Omega=9/2,7/2,...,1/2$ stemming from the $1g_{9/2}$ and the negative 
parity 
orbitals with $\Omega=3/2,1/2$ arising from the $2p_{3/2}$ 
(those just below the Fermi surface, See Fig.4).
Among the possible couplings associated with the octupole 
deformations,
the $\Delta\Omega=3$ coupling between the
positive parity $\Omega=9/2$  and negative parity
$\Omega=3/2$ orbitals, and also the one between the 
positive parity $\Omega=7/2$ and negative parity
$\Omega=1/2$ orbitals have the smallest energy difference, 
and give enhanced softness toward the
triangular $Y_{33}$ deformation.

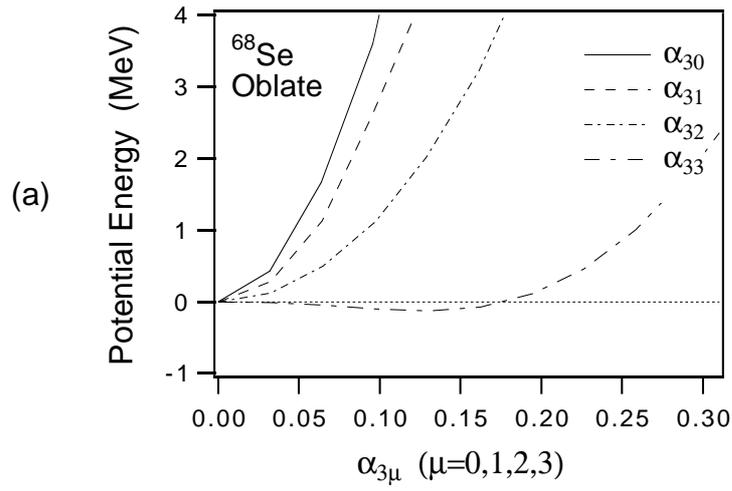
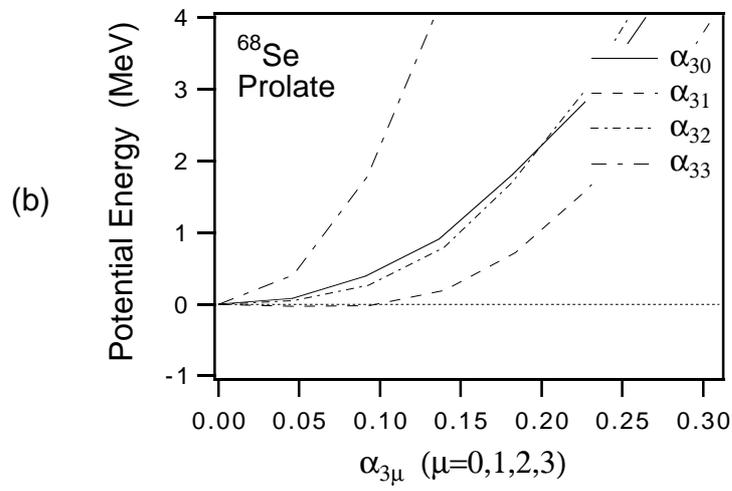
\begin{figure}
\caption{Potential energy surfaces with respect to the different
types of octupole deformations, calculated for the oblate ground state (a) 
and the second minimum prolate state (b) of $^{68}$Se.
The quadrupole deformations are set to $\beta=0.25, \gamma=60^{\circ}$ 
and $\beta=0.25, \gamma=0^{\circ}$ for the ground and second minimum states, 
respectively.}
\label{figure3}
\end{figure}

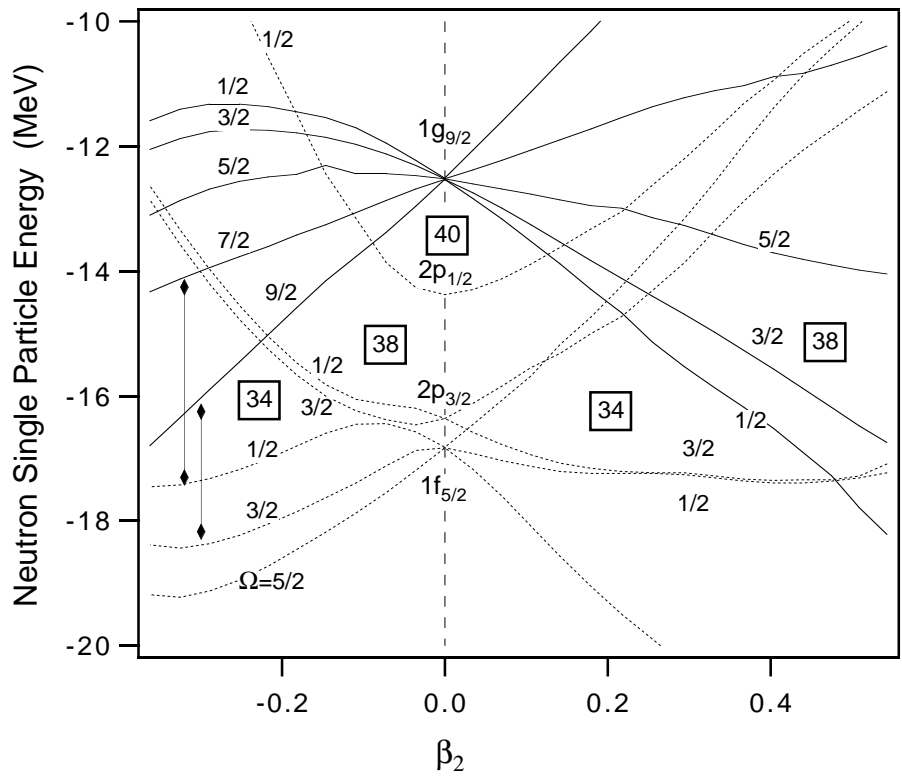
\begin{figure}
\caption{The neutron single-particle  
levels for $^{80}$Zr as a function 
of the quadrupole deformation parameter $\beta_{2}$ calculated with
the quadrupole constraint and the axial
and reflection symmetries.
For each orbitals, we put the value of $\Omega$, the projection
of the angular momentum along the symmetry axis.
The arrows indicate the $\Delta\Omega=3$ coupling associated
with the triangular $Y_{33}$ deformation
 as discussed in the text.
}
\label{figure4}
\end{figure}

In the prolate solutions of $^{68}$Se and $^{72}$Kr, 
the non-axial octupole deformations with finite
value of $\beta_{31}$ (so-called 'banana' shape) are also found
although the magnitude of the deformation is small.
The associated potential energy surface for the octupole deformations,
shown in Fig.\ref{figure3}.(b), indicates that the $\alpha_{31}$
direction is  soft while the other types including $\alpha_{30}$
pear-like deformation are not. It is noted that the
positive parity orbital with $\Omega=3/2$ (stemming from $1g_{9/2}$)
and the negative parity  $\Omega=1/2$ ($2p_{3/2}1f_{1/2}$) located
near the $N,Z=34-36$ Fermi surface (see Fig.4) are responsible for the
$Y_{31}$ deformation. Although there exists closely-lying
 $\Delta\Omega=0$ pair of 
 $\Omega=1/2$ ($1g_{9/2}$)  and $\Omega=1/2$
($2p_{3/2}1f_{1/2}$) orbitals, the matrix element of
the $\Delta\Omega=0$ octupole operator $r^3Y_{30}$ between
these orbitals is expected to be small, provided these
orbitals have good asymptotic Nilsson
[440]$\frac{1}{2}$ and [310]$\frac{1}{2}$ components, respectively.

In summary, the self-consistent mean-field calculation
predicts that $^{80}$Zr has the low-lying coexisting state
showing the spontaneous  $Y_{32}$ tetrahedral deformation and
the oblate ground state in $^{68}$Se is extremely soft with respect 
to the $Y_{33}$ triangular deformation. 
Appearances of the non-axial octupole
deformations are governed by the single-particle shell structures
at $N,Z \sim 40$ and $N,Z \sim 34$ for the tetrahedral and 
triangular deformations, respectively.

The non-axial octupole deformations discussed above 
will accompany specific spectroscopic properties of excited states.
The intrinsic tetrahedron shape having the $T_{d}$
symmetry leads to a series of levels with
$I^{\pi}=0^{+},3^{-},4^{+},6^{+},7^{-}, \cdots$, 
which follow the  $E(I)-E(0) \sim I(I+1)$ relation and 
are connected with strongly enhanced E3 transitions\cite{TETRA}.
In $^{80}$Zr, for example, a low-lying excited $J^{\pi}=3^{-}$ 
state is expected, and one could observe a strong E3 decay 
from a $3^-$ state to a low-lying $0^{+}$ state  
associated with the tetrahedral deformed second 
minimum.
The oblate deformed ground state of $^{68}$Se which is extremely 
soft towards the $Y_{33}$ direction will accompany a low-lying 
$3^{-}$ excited state with high collectivity.
Furthermore, the intrinsic triangular shape having the $D_{3h}$
symmetry  may form the 
characteristic band structure, that is, 
the ground band $I^{+}=0^{+},2^{+},4^{+},\cdots$ and 
a $K^\pi=3^-$ band $I^{-}=3^{-},4^{-},5^{-},\cdots$
as seen in $^{12}$C \cite{CLUSTER}.

The authors would like to thank Prof. K. Matsuyanagi for helpful discussions.
Numerical computation in this work was mainly performed 
at the Yukawa Institute Computer Facility.
Part of numerical calculations were also performed 
on the FACOM VPP-500 supercomputer in RIKEN.

\begin{table}
\begin{center}
\begin{tabular}{c|ccccc}
        \hline \hline
	 & \multicolumn{2}{c}{Oblate} & Spherical 
	 & \multicolumn{2}{c}{Prolate}\\ \hline \hline

	 & & g.s.  &     & 0.62 & 4.00 \\
 $^{64}$Ge
	 &
 	 & $\beta ,\gamma = 0.27, 25^{\circ}$ (T) 
 	 & 
	 & $\beta ,\gamma = 0.24, 6^{\circ}$ 
	 & $\beta ,\gamma = 0.38, 0^{\circ}$ \\

	 &
	 &  $\beta_{3} = \beta_{33}=0.01$ 
	 &
	 &  $\beta_{3} = 0.00$
	 &  $\beta_{3} = 0.00$ \\ \hline

	 & &  g.s. &   & 0.32 & 2.42 \\
 $^{68}$Se
	 &
	 & $\beta ,\gamma  = 0.25, 60^{\circ}$ 
	 & 
	 & $\beta ,\gamma = 0.25, 0^{\circ}$ 
	 & $\beta ,\gamma = 0.40, 18^{\circ}$ (T)\\

	 &
	 &  $\beta_{3} = \beta_{33}=0.15$ 
	 &  
	 &  $\beta_{3} = \beta_{31}=0.06$ 
	 &  $\beta_{3} = \beta_{31}=0.02$ \\ \hline

	 & g.s. &  1.12 & & 1.74 & \\
 $^{72}$Kr
	 & $\beta ,\gamma = 0.34, 60^{\circ}$
	 & $\beta ,\gamma  = 0.27, 58^{\circ}$ 
	 & 
	 & $\beta ,\gamma = 0.42, 1^{\circ}$ 
	 & \\

	 &  $\beta_{3} = 0.00$
	 &  $\beta_{3} = \beta_{33}=0.05$ 
	 &  
	 &  $\beta_{3} = \beta_{31}=0.03$ 
	 & \\ \hline

	 & & 2.58 & 3.25 & & g.s. \\
 $^{76}$Sr
	 &
	 & $\beta ,\gamma  = 0.13, 60^{\circ}$ 
	 & $\beta ,\gamma  = 0.02, 0^{\circ}$ 
	 &  
	 & $\beta ,\gamma = 0.49, 0^{\circ}$ \\

	 &
	 &  $\beta_{3} = \beta_{33}=0.16$
	 &  $\beta_{3} = \beta_{32}=0.12$
	 &  
	 &  $\beta_{3} = 0.00$ \\ \hline

	 & &  1.58 & 0.90 & & g.s. \\
 $^{80}$Zr
	 &
	 & $\beta ,\gamma  = 0.20, 59^{\circ}$ 
	 & $\beta ,\gamma = 0.00, 0^{\circ}$
	 & 
	 & $\beta ,\gamma = 0.50, 0^{\circ}$ \\
 
	 &
	 &  $\beta_{3} = \beta_{32}=0.04$ 
	 &  $\beta_{3} = \beta_{32}=0.24$
	 &  
	 &  $\beta_{3} = 0.00$ \\ \hline

	 & &  g.s. & 0.24 &  & 0.85 \\
 $^{84}$Mo
	 &
	 & $\beta ,\gamma  = 0.20, 56^{\circ}$ 
	 & $\beta ,\gamma = 0.05, 60^{\circ}$
	 & 
	 & $\beta ,\gamma = 0.64, 0^{\circ}$ \\
 
	 &
	 &  $\beta_{3} = 0.00$ 
	 &  $\beta_{3} = \beta_{30} = 0.13$
	 &  
	 &  $\beta_{3} = 0.00$ \\ \hline \hline
\end{tabular}
\caption{The ground states and the local minimum states obtained 
in the present SHF+BCS calculation. The energy difference (MeV) 
between the ground state and the local minimum state (the ground 
state is denoted as g.s.), the quadrupole and octupole deformation 
parameters are shown, where label (T) stands for the triaxial 
deformation. For the solutions showing the octupole deformation, 
the most dominant component of $\beta_{3m}$ is also
presented.  
Each solutions are classified into the three groups, oblate, 
spherical and prolate  by their quadrupole deformation parameter, 
except for the ground state of $^{64}$Ge. 
The ground state of $^{64}$Ge which shows the triaxial 
deformation is classified into the oblate group.}
\label{table1}
\end{center}
\end{table}

\end{document}